\documentclass{appolb}
\usepackage{graphicx}
\usepackage{amsmath}

\def\vec#1{\mathbf{#1}}

\begin{document} 
\pagestyle{plain}
\title{Kinetic description of fermion production in the oscillator representation 
}
\author{V.N.~Pervushin, and V.V.~Skokov 
\address{Bogoliubov Laboratory of Theoretical Physics,\\
Joint Institute for Nuclear Research,
141980, Dubna, Russia}}
\maketitle
\begin{abstract}
We investigate the fermion creation  in quantum kinetic theory by 
applying ``oscillator representation'' approach, which was 
earlier developed  for bosonic systems. 
We show that
in some particular cases
(Yukawa-like interaction, fixed direction of 
external vector field) resulting  Kinetic Equation (KE)  
reduces to KE obtained by time-dependent Bogoliubov transformation method. 
We conclude ``oscillator representation'' approach to be more universal for the derivation
of quantum transport equations  in strong space-homogeneous time-dependent fields. 
We discuss some possible applications of obtained
KE to cosmology and particle production  in strong laser fields. 
\end{abstract}

\section{Introduction}
Spontaneous particle production of pairs under
the action of strong external fields attracts a lot of attention 
since it is  a multi-particle non-perturbative effect which is 
still lacking experimental verification. According to the famous 
Schwinger formula~\cite{Schwinger:1951nm} the probability of particle creation 
in the constant electromagnetic field becomes  essential 
if the field reaches the critical value $E^c=m^2/e$. Considering 
electron-positron pair creation in laser field we may conclude that 
one must 
have unattainable static field $E^c_{e^+ e^-} = 1.32 \cdot 10^{16}$ V/cm. 
However theoretical and numerical results obtained  in recent papers
(e.g. \cite{Popov:2001ak,Ringwald:2001ib,Prozorkevich:2004yp,Blaschke:2004hb,Blaschke:2005hs}) 
show that the probability of particle creation in {\em time-dependent} fields
can be gradually enhanced in contrast to static ones.  
Also recent developments in laser technology (method of chirped pulse amplification~\cite{BulanovTajima})
and underway construction of X-ray free electron lasers  will allow  an experimental 
 verification of spontaneous particle creation in the nearest future. 

Fermion production in external time-dependent fields 
was considered in a number of papers \cite{Marinov:1977gq,grib80,cooper12,Smolyansky:1997fc}. 
The method of the time-dependent Bogoliubov transformation\footnote{The Bogoliubov transformations 
are named after N.N. Bogoliubov who introduced and applied these canonical 
transformations to determine integrals of motion for  creation and annihilation operators~\cite{Bogoliubov}.}
used there faces difficulties 
if the external field has more than one non-zero component, thus KE was obtained
only for the linear polarized   (also  called  flux-tube  geometry)  
field ($A_\mu = (0, 0, 0, A(t) = A_3(t))$). 
As was shown earlier for boson creation~\cite{holomorphic} in periodic 
laser field, pair production is efficient if the field has rather the circular 
polarisation than the linear one\footnote{As was noted in~\cite{Blaschke:2004hb}
the linear polarised field is not appropriate for a quantitative description of the laser pulse. 
 However,the linear polarised field was considered~\cite{Popov:2001ak,Blaschke:2004hb,Blaschke:2005hs,laser2}
 to obtain   qualitative level results.}.

The approach presented 
in current paper
resolves the problem  of fermion creation in an arbitrary polarised field (e.g. 
in Hamilton gauge $A_\mu = (0, A_1(t), A_2(t),  A_3(t))$ ).
This result can be  of great importance for the theoretical 
predictions about dilepton and photon  yield in  high-intensity 
laser 
experiments~\cite{laser2, laser1}.

As  already mentioned, the classical approach to  the description of the particle creation in the 
external fields or in curved space  is based on the  time-dependent 
Bogoliubov  transformation~\cite{Marinov:1977gq,grib80}.  Concentrating on fermion fields we  will briefly 
summarise this approach. First of all let us suppose that in the asymptotic states $t \to \pm \infty$
the external field vanishes. It means that for the limit  $t \to \pm \infty$
the interpretation in terms  of particles and antiparticles  becomes possible. Thus 
we can introduce the orthogonal set of  functions $\psi^{\pm}_{k s}(x)$, which 
represent the positive and negative frequency solution at  $t \to - \infty$. The field function can 
 be decomposed as
\begin{equation}
\psi (x) =   \sum_{s}  \int\frac{d^3k}{(2 \pi)^{3/2}}\left( \psi^+_{s}(x;\vec{k}) a_{s}(\vec{k}) 
+   \psi^+_{s}(x;\vec{k}) b_s^+(\vec{k}) \right), 
\label{class}
\end{equation}
where we perform the sum over spin indices $s$. $a_{s}$($b_s^+$) is the annihilation (creation) operator
of the (anti)particle with the spin $s$ and momentum $\vec{k}$. 
If we demand that the creation and annihilation operators introduced in this way satisfy general anticommutational
relations, then the Fock space can be  constructed with the vacuum state defined at  $t \to - \infty$.
However, following this 
procedure  one can easily prove that Hamiltonian in the external time-dependent  field will be non-diagonal
(except  the initial  state  $t \to - \infty$). To diagonalize  it one can use Bogoliubov time-dependent 
transformation  to redefine the particle creation  $c_{s}(t)$  and annihilation  $d_{s}^+(t)$ operators 
with nontrivial equations of motion for them. 
The  particle number  in this formalism can be defined at every time moment as 
$N_{\vec{k} s}(t) \equiv f_s (\vec{k},t) = \langle0\vert c^+_{s}(\vec{k};t) c_{s}(\vec{k};t)    \vert 0 \rangle$.

In this work, we will use another approach --- the so-called ``oscillator representation'', which was earlier
developed for bosonic systems~\cite{holomorphic,ps}. The main idea is to 
construct the field function and to define the time-dependent  creation and annihilation operators 
so that the Hamiltonian immediately becomes   diagonal.  Moreover it  
should take the form of  the free field Hamiltonian  at  $t \to - \infty$. 
The solution of equations of motion derived  from Dirac equation   for introduced operators 
will provide   the wanted distribution function $f_s (\vec{k},t)$.

This paper is organised  as follows. 
In Section 2 
the oscillator  representation formalism for fermions is given. In Section 3   
the kinetic equation is derived.  
The  applications of the model
and particularly  the fermion production in the early Universe are  considered in Section 4.  
Section  5 summarises the article. 
Appendix contains  the KE in the convenient form for numerical calculations.

\section{Oscillator (Holomorphic) Representation}

We start from the Lagrangian for electrodynamics 
\begin{eqnarray}%
\label{lag}%
 L =  \bar \psi ( i D_\mu\gamma^\mu - m(t)) \psi,
\end{eqnarray}%
%
where $D_\alpha$ is the covariant  derivative %
\begin{eqnarray}%
D_\alpha = \partial_\alpha + i e A_\alpha.%
\end{eqnarray}%

Let us  assume that the 
function  $m(t)$ is an  arbitrary  smooth function of time.  
We also demand  that the  electric field  is  classical and space-homogeneous with
a  continuous  $A_\alpha(t)$.  
The classical approximation is  supported by  the leading
order of a large $N$ approximation, see e.g. \cite{cooper12}.

In order to derive the correct oscillator representation of the fermion field 
one can use  the Foldy-Wouthuysen transformation~\cite{Itzykson}:
\begin{eqnarray}
 \label{OR_fermions:psi}
  \psi (x) &=&  \int \frac{d^3p}{(2 \pi)^{3/2}} e^{i\vec{p}\vec{x}}
 e^{iS(\vec{p},t)} \left( a_s(\vec{p},t) u_s(0)   +  b_s^+
 (-\vec{p},t) v_s(0)  \right), \\
 \label{OR_fermions:psihat}
 \bar{\psi} (x) &=&  \int \frac{d^3p}{(2 \pi)^{3/2}} e^{-i\vec{p}\vec{x}}
  \left( a^+_s(\vec{p},t) \bar{u}_s(0)   +  b_s
 (-\vec{p},t) \bar{v_s}(0)  \right) {e^{iS(\vec{p},t)}},
\end{eqnarray}
where summation over spinor indices ($s=1,2$) is implied, and spinors
$u(0), v(0)$ are defined  as a solution of free Dirac equation in the particle rest frame:
\begin{eqnarray}
 \label{OR_fermions:zs}
 (\gamma_0-1)u(0) &=& 0,\\
 (\gamma_0+1)v(0) &=& 0.
\end{eqnarray}

The unitary operator  $e^{iS(\vec{p},t)}$   can be dependent both on time and momentum.  
According to the scheme of oscillator  representation   $e^{iS(\vec{p},t)}$ is defined  so that 
the Hamiltonian density
after the field decomposition
\begin{eqnarray}
H(\vec{p},t) &=& \left( a^+_s(\vec{p},t) \bar{u}_s(0)   +  b_s
 (-\vec{p},t) \bar{v_s}(0)  \right) \times \nonumber  \\
 &&{e^{i S(\vec{p},t)}} (P_i \gamma_i + m) e^{iS(\vec{p},t)} \times \nonumber \\
 &&\left( a_s(\vec{p},t) u_s(0)   +  b_s^+
 (-\vec{p},t) v_s(0)  \right)
\label{OR_fermions:Hamilt}
\end{eqnarray}
becomes  diagonal, i.e.  
\begin{eqnarray}
H(\vec{p},t) &=& \left( a^+_s(\vec{p},t) \bar{u}_s(0)   +  b_s
 (-\vec{p},t) \bar{v_s}(0)  \right) \times \nonumber  \\
 &&\omega(\vec{P},t) \times \left( a_s(\vec{p},t) u_s(0)   +  b_s^+
 (-\vec{p},t) v_s(0)  \right), 
\label{OR_fermions:sHamilt}
\end{eqnarray}
with  quasiparticle energy   $\omega(\vec{P},t) = \sqrt{\vec{P}^2+m^2}$ defined 
by   kinetic momentum $\vec{P} = \vec{p} - e \vec{A} $.

This condition can be rewritten as
\begin{equation}
e^{iS(\vec{p},t)} (P_i \gamma_i + m)e^{i S(\vec{p},t)} =
\omega(\vec{P},t),
\label{OR_fermions:cond}
\end{equation}
with the following solution
\begin{equation}
\label{OR_fermions:SolutS}
e^{iS(\vec{p},t)} = \frac{\omega(\vec{P},t)+ m  + P_i \gamma^i
}{\sqrt{2\omega(\vec{P},t)(m+\omega(\vec{P},t))}}.
\end{equation}
One can write down the  explicit expression  for $S$ in a close form
\begin{equation}
\label{OR_fermions:Un}
S(\vec{p},t) = i \frac{P_i \gamma_i}{|\vec{P}|} \phi,
\end{equation}
where $\phi$ is given by
\begin{equation}
\label{OR_fermions:Angle}
 \sin (2 \phi)  = \frac{|\vec{P}|}{\omega(\vec{P},t)},\,\,\, \cos (2 \phi)=
 \frac{m}{\omega(\vec{P},t)}.
\end{equation}

It is easy to convince  that
\begin{eqnarray}
\label{OR_fermions:u}
u(\vec{p} \to \vec{P}) &=& e^{iS(\vec{p},t)}  u(0),\\
v(-(\vec{p} \to \vec{P}) ) &=& e^{iS(\vec{p},t)}  v(0),
\label{OR_fermions:v}
\end{eqnarray}
where $u(\vec{p}), v(\vec{p})$ are the solutions of free Dirac
equation.
Taking into account  these  expressions (\ref{OR_fermions:u}), (\ref{OR_fermions:v})  we 
can rewrite 
the  decomposition of Eqs. (\ref{OR_fermions:psi}), (\ref{OR_fermions:psihat}) 
in the form 
\begin{eqnarray}
 \psi (x) &=&  \int \frac{d^3p}{(2 \pi)^{3/2}} e^{i\vec{p}\vec{x}}
 \left\{ a_s(\vec{p},t)
  u_s(\vec{P})   +  b_s^+
 (-\vec{p},t)  v_s(-\vec{P})  \right\}, \\
 \bar{\psi}(x) &=&  \int \frac{d^3p}{(2 \pi)^{3/2}} e^{-i\vec{p}\vec{x}} \left\{
 a_s^+(\vec{p},t)
 \bar{ u}_s(\vec{P})  +  b_s
 (-\vec{p},t) \bar{ v}_s(-\vec{P})  \right\}.
\end{eqnarray}
Thus, the formal substitution $\vec{p} \to \vec{P}$ in the free field
decomposition results in  (\ref{OR_fermions:psi}), (\ref{OR_fermions:psihat}).

The equations of motion for the creation and annihilation operators
can be obtained from Dirac equations (dots denote the derivative with respect to time):
\begin{eqnarray}
\dot{a}_{s'}(\vec{p},t) &=& -i \omega a_{s'}(\vec{p},t) -  \Theta^u_{s's}  a_s(\vec{p},t) -
 \Xi^u_{s's} b^+_s(-\vec{p},t),\\
\dot{b}^+_{s'}(-\vec{p},t) &=& i \omega   b^+_{s'}(-\vec{p},t) -  \Theta^v_{s's}  b^+_{s}(-\vec{p},t) -
 \Xi^v_{s's}  a_s(\vec{p},t) ,
\end{eqnarray}
or in matrix form 
\begin{eqnarray}
\label{m1}
\dot{a}(\vec{p},t) &=& -i \omega a(\vec{p},t) - \Theta^u a(\vec{p},t) - \Xi^u b^+(-\vec{p},t),\\
\label{m2}
\dot{b}^+(-\vec{p},t) &=& i \omega   b^+(-\vec{p},t) - \Theta^v b^+(-\vec{p},t) - \Xi^v  a(\vec{p},t), 
\end{eqnarray}
where $\Theta^{u,v}$, $\Xi^{u,v}$  are given by  
\begin{eqnarray}
\Theta^u_{s' s} &=& u^+_{s'}(0) e^{-iS} \partial_0 e^{iS} u_s(0),\\
\Theta^v_{s' s} &=& v^+_{s'}(0) e^{-iS} \partial_0 e^{iS} v_s(0),\\
\Xi^u_{s' s} &=& u^+_{s'}(0) e^{-iS} \partial_0 e^{iS} v_s(0),\\
\Xi^v_{s' s} &=& v^+_{s'}(0) e^{-iS} \partial_0 e^{iS} u_s(0).
\label{matrix_not}
\end{eqnarray}
The straightforward calculation  gives
\begin{eqnarray}
\Theta^u &=& \Theta^v = i e  \frac{[\vec{P}\times\vec{E}]}{2 \omega(\omega +m)} \mbox{\boldmath $\sigma$},  \\
\Xi^u &=& - \Xi^v =  \frac{ \dot{\omega}+\dot{m}}{2 \omega(\omega+m)} \vec{P} \mbox{\boldmath $\sigma$} - 
\frac{e}{2 \omega} \vec{E} \mbox{\boldmath $\sigma$}, 
\label{matrix_explicit}
\end{eqnarray}
where  $\sigma_k$ are  Pauli matrices and $\vec{E} = - \dot{\vec{A}}$ is the field strength. 

In  infinite past $t \to -  \infty$,   the operators  $a_{s}$ and $b_{s}$
should satisfy the classical  anticommutation relations. 
However one can see that the second terms in the right part of 
the equations of motion (\ref{m1}, \ref{m2})  can  modify    
the classical  anticommutation relations in the presence of the time-dependent field. 
Indeed  let us assume that  
\begin{equation}
\label{OR_fermions:Comm}
 \{a_s(\vec{k},t), a^+_{s'}(\vec{k'},t) \} =
 \Pi_{ss'}(\vec{k},t)\delta^3(\vec{k}-\vec{k'}), 
\end{equation}
where the equation for unknown time-dependent  matrix $\Pi_{ss'}$ 
is derived  by using  (\ref{m1}, \ref{m2})  
\begin{equation}
\dot{\Pi}(\vec{k},t) =  \left[ \Pi(\vec{k},t),  \Theta^u(\vec{k},t) \right].  
\label{OR_fermions:Pi}
\end{equation}
However, taking into account trivial symmetries of the matrix  $\Pi_{ss'}$ 
\begin{equation}
\Pi_{ss'} = \Pi_{s's}=\Pi^*_{s's} 
\end{equation}
and initial conditions in infinite  past $\Pi_{ss'}(\vec{k},t \to -  \infty) = \delta_{ss'}$
one easily  proves that  $\Pi_{ss'}(t) = {\rm const}  =  \delta_{ss'}$, i.e. canonical 
commutation relations are valid even  in time-dependent external  field.

\section{Distribution Function and Kinetic Equation}

In this section we will derive the equation of motion (or KE)  for the one-particle correlator
\begin{equation}
\label{Ms}%
f_{s s'}(\vec{p},t) = \langle0|a_{s'}^+(\vec{p},t) a_s(\vec{p},t) |0\rangle.  
\end{equation}
The distribution function is the diagonal components of $f_{s s'}(\vec{p},t)$~\cite{grib80,cooper12,QGP1}. 
Further calculations will show that $f_{s s'}$ is coupled to one-particle 
correlator describing the antiparticle distribution $g_{s s'}$ and two 
anomalous correlators $y^\pm_{s s'}$:
\begin{eqnarray}
\label{OPC_beg}
 g_{s s'}(\vec{p},t) &=& \langle 0| b_{s'}(-\vec{p},t) b^+_s(-\vec{p},t)  |0\rangle, \\
y^-_{s s'}( \vec{p},t)  &=& \langle 0| b_{s'}(-\vec{p},t) a_s( \vec{p},t) |0\rangle,   \\
y^+_{s s'}( \vec{p},t)  &=& \langle 0|  a^+_{s'}( \vec{p},t) b^+_s(-\vec{p},t)  |0\rangle.
\label{OPC_end}
\end{eqnarray}

KE is obtained by  differentiating  (\ref{Ms}-\ref{OPC_end})  with respect to time  and subsequent 
substituting the equations of motion~(\ref{m1}-\ref{m2}) and definitions (\ref{Ms}-\ref{OPC_end}). 
After some trivial transformation one gets:
\begin{eqnarray}
\label{not_final_yet_beg}
 \dot{f} &=& i [ f , \Theta ]  -  y^- \Xi - \Xi y^+,\\
 \dot{y}^- &=& - 2 i \omega y^-  + i [ y^- , \Theta] + f \Xi  -\Xi g,\\
 \dot{y}^+ &=& 2 i \omega y^+ + i [y^+, \Theta]  - g \Xi + \Xi f, \\
 \dot{g} &=& i [ g , \Theta ]  +  y^- \Xi + \Xi y^+,
\label{not_final_yet}
\end{eqnarray}
where $\Theta = \Theta^u/i = \Theta_i \sigma_i$ and $\Xi = \Xi^u = \Xi_i \sigma_i$ are hermitian
matrices.  
It is more convenient to introduce two hermitian matrices
\begin{eqnarray}
r^+ &=& \frac{1}{2}( y^+ + y^-), \\
r^- &=& \frac{i}{2}( y^+ - y^-)
\label{Upsilon}
\end{eqnarray}
and decompose  $\chi= f,\ g,\ r^\pm$ into U(2) basis ($I$ is a unitary matrix)
\begin{eqnarray}
\label{matr_decomp1}
\chi  &=& \chi_0 I  + \chi_i \sigma_i,\,\, i = 1,2,3;\\
\chi_0 &=& \frac{1}{2} \mbox{Tr} \chi, \, \, \chi_i = \frac{1}{2} \mbox{Tr} (\chi \sigma_i).
\label{matr_decomp}
\end{eqnarray}
Finally in terms of (\ref{matr_decomp1}-\ref{matr_decomp}) KE (\ref{not_final_yet_beg}-\ref{not_final_yet})
takes the following form:
\begin{eqnarray}
\label{general_final_beg}
 \dot{f}_0 &=& - 2  \vec{\Xi} \vec{r}^+, \\
\label{general_final_beg1}
 \dot{\vec{f}} &=& - 2 [\vec{f} \times \vec{\Theta}] + 2  [\vec{r}^- \times \vec{\Xi}]
 - 2 r^+_0 \vec{\Xi},\\
\label{general_final_beg2}
 \dot{g}_0 &=&  2 \vec{\Xi} \vec{r}^+, \\
\label{general_final_beg3}
 \dot{\vec{g}} &=& - 2 [\vec{g} \times \vec{\Theta}] - 2  [\vec{r}^- \times \vec{\Xi}]
+ 2 r^+_0 \vec{\Xi}, \\
\label{general_final_beg4}
 \dot{r}_0^+ &=& 2 \omega r^-_0 + \vec{\Xi}  \vec{f} - \vec{\Xi}  \vec{g},\\
 \dot{\vec{r}}^+ &=& 2 \omega \vec{r}^- - 2 [\vec{r}^+ \times \vec{\Theta}]
+\vec{\Xi} (f_0 - g_0), \\
 \dot{r}_0^- &=& - 2 \omega r^+_0,\\
 \dot{\vec{r}}^- &=& -2 \omega \vec{r}^+ - 2 [\vec{r}^- \times \vec{\Theta}]
+ [\vec{f} \times \vec{\Xi}] +  [\vec{g} \times \vec{\Xi}].
\label{general_final}
\end{eqnarray}

The KE (\ref{general_final_beg}-\ref{general_final}) allows some further simplifications
if we assume zero initial conditions in infinite past $t\to-\infty$  for $f_0, \vec{f} =0 $ (absence of particles), 
$r^\pm_0, \vec{r}^\pm = 0$ (anomalous correlators turns to zero at in-vacuum), and $g_0=1, \vec{g}=0$
(neutrality condition and absence of antiparticles). Summing (\ref{general_final_beg}) and (\ref{general_final_beg2}), 
 (\ref{general_final_beg1}) and (\ref{general_final_beg3})  one finds the following integrals of motion:
\begin{eqnarray}
f_0&=&1-g_0,\\
\vec{f}&=&-\vec{g}.
\end{eqnarray}
Taking them into account KE equation (\ref{general_final_beg}-\ref{general_final}) is rewritten as 
\begin{eqnarray}
\label{general_final_beg_}
 \dot{f}_0 &=& - 2  \vec{\Xi} \vec{r}^+, \\
\label{general_final_beg1_}
 \dot{\vec{f}} &=& - 2 [\vec{f} \times \vec{\Theta}] + 2  [\vec{r}^- \times \vec{\Xi}]
 - 2 r^+_0 \vec{\Xi},\\
\label{general_final_beg2_}
 \dot{r}_0^+ &=& 2 \omega r^-_0 + 2 \vec{\Xi}  \vec{f},\\
 \dot{\vec{r}}^+ &=& 2 \omega \vec{r}^- - 2 [\vec{r}^+ \times \vec{\Theta}]
+\vec{\Xi} (2 f_0 - 1), \\
 \dot{r}_0^- &=& - 2 \omega r^+_0,\\
 \dot{\vec{r}}^- &=& -2 \omega \vec{r}^+ - 2 [\vec{r}^- \times \vec{\Theta}].
\label{general_final_}
\end{eqnarray}

\section{Applications}

\subsection{Scalar-like interaction}
If the vector field vanishes  $A_i(t) = 0, E_i(t) = 0$, 
we obtain 
\begin{eqnarray}
\vec{\Theta} &=& 0, \\
\vec{\Xi} &=& \frac{\dot{m}}{2 \omega^2}  \vec{p}.  
\label{cosm_xi}
\end{eqnarray}

For this case the KE (\ref{general_final_beg}-\ref{general_final}) allows  some  simplification. 
First of all, the vector components $\vec{r}^{\pm}$, $\vec{f}$,  $\vec{g}$ will be collinear 
to $\vec{p}$. Thus let us  redefine 
\begin{equation}
\vec{r}^{\pm} = \mp v^{\pm} \frac{\vec{p}}{|\vec{p}|} 
\label{redefinition}
\end{equation}
and assuming zero initial conditions in the asymptotic past rewrite the KE in the form  
\begin{eqnarray}
\label{KE_mass_time_dependent_begin}
\dot{f}_0 &=&  2 W v^+, \\
\dot{v}^+ &=&  W (1 - 2 f_0) -  2 \omega v^-, \\
\dot{v}^- &=&  2 \omega v^+, 
\label{KE_mass_time_dependent}
\end{eqnarray}
where $W = \frac{\dot{m}}{2 \omega^2}  |\vec{p}|$ and in the above equations the 
neutrality condition $g_0=1-f_0$ was used.  
Applying zero initial conditions to $f_0$ and $\vec{f}$  we can conclude that the
vector components of the distribution function for fermions
and anti-fermions are identical  zero and  the evolution  of $f_0$ component is 
given by (\ref{KE_mass_time_dependent_begin}-\ref{KE_mass_time_dependent}).

So far we have not discussed any origin of the time dependence of the  mass $m(t)$. 
However any fermion interaction of the Yukawa type $\Delta L = g \phi \bar{\psi} \psi$ 
in the mean field approximation for the scalar field $\phi$  results 
in  the time-dependent effective fermion mass $m(t) = m + g\,   \langle \phi \rangle$.
Such a type of interactions is presented  in the Higgs model,  sigma model and can 
appear in some  effective field theories (e.g. Walecka model~\cite{Walecka:1974qa}). 

Here we will concentrate on  another important example where time-dependence of the effective 
fermion mass can be generated, namely, conformal field theory. 
\subsubsection{Fermion production in the Early Universe}
For the sake of consistency let us recall some general and well-known 
facts of general relativity. 

The Lagrangian of the massive fermion field in general relativity is written as
\begin{equation}
L = \sqrt{-g} \left[i \bar{\psi}(x)  \gamma^n(x) \Delta_n \psi(x) -  m  \bar{\psi}(x) \psi(x) \right], 
\label{GR_L}
\end{equation}
where 
\begin{eqnarray}
\gamma^n(x) &=&  h^n_{(\alpha)} \gamma^\alpha,\\
\Delta_n &=& \partial_n + \frac{1}{4} C_{\mu \nu \rho} h_n^{(\rho)} \gamma^\nu \gamma^\mu,   
\label{def}
\end{eqnarray}
The rotation coefficients can be defined  by means of tetrad  as (for details, refer to~\cite{grib80,Birrel82}) 
\begin{equation}
C_{\mu \nu \rho} = (\Delta_m h^n_{(\mu)}) h_{(\nu)n} h^m_{(\rho)}.  
\end{equation}

We will consider conformally static space-time. In this case the metric is conformally 
related to a static space-time. For a particular case of Robertson-Walker metric we have
\begin{eqnarray}
g_{\mu\nu} &=& a^2(\eta) h_{\mu\nu}, \\
ds^2 &=& a^2(\eta) (d \eta^2 - d \vec{x}^2),
\label{metric}
\end{eqnarray}
where $a$ is the conformal factor, $\eta = \int dt \ a(t)$  is the conformal time, $h_{\mu \nu}$ is the
Minkowski metric.  Introducing new field variables $\psi_c = a^{1/2}(\eta) \psi$ we obtain
\begin{equation}
L = a^2(\eta) \left[ i \bar{\psi}_c(x)  \gamma^\mu \partial_\mu \psi_c(x) -  m(\eta)  \bar{\psi}_c(x) \psi_c(x) \right], 
\label{C_L}
\end{equation}
where we define the effective fermion mass $m(\eta) \equiv m a(\eta)$.  

\begin{figure}[t]
\centerline{ \includegraphics[width=7cm]{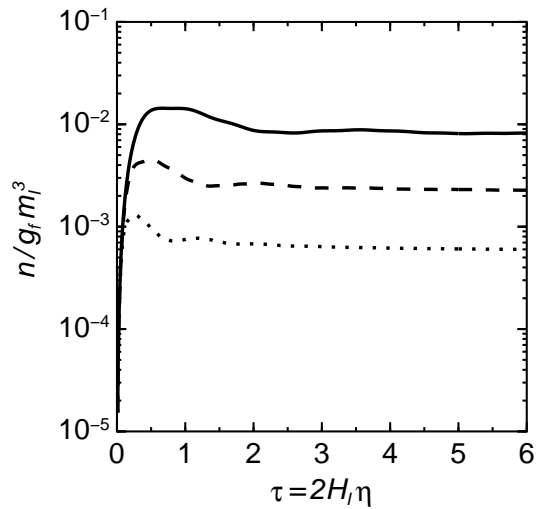}}
 \caption{The time dependence of the 
 number density  for  fermions with  different masses $\frac{m_{I}}{2 H_I}$ = 0.5, 1, 2 (from top to bottom).}
\end{figure}
\begin{figure}[t]
\centerline{ \includegraphics[width=7cm]{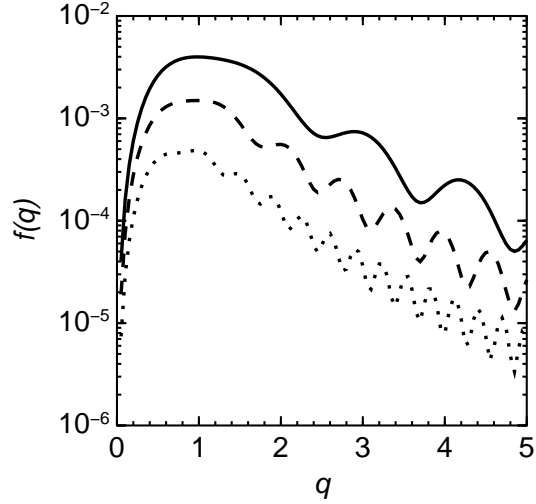}}
 \caption{The momentum dependence of the distribution function for fermions with different  masses 
 $\frac{m_{I}}{2 H_I}$ = 0.5, 1, 2 (from top to bottom) at the fixed moment of time $\tau \equiv 2 H_I t = 6$; 
 $q \equiv \vert \vec{q}\vert$ is dimensionless momentum $\vec{q} = \vec{p}/m_I$.}
\end{figure}
\begin{figure}[b]
\centerline{ \includegraphics[width=7cm]{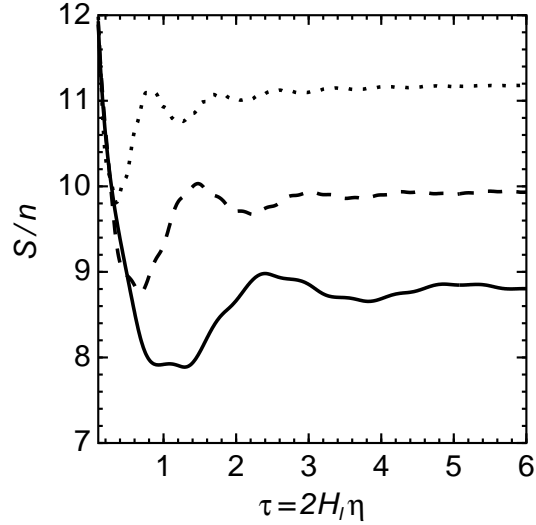}}
 \caption{The time dependence of the specific entropy $S/n$~(\ref{ent}) for fermions
 with different   masses 
 $\frac{m_{I}}{2 H_I}$ = 0.5, 1, 2 (from bottom to top). }
\end{figure}

The cosmological creation of particles is considered  in terms of 
the so-called conformal Universe~\cite{Pervushin3}, where
  the volume of the Universe does not increase, while
all masses, including the Planck mass, are scaled by the cosmic factor $a(\eta)$.
We assume a stiff state  scenario, where the mass $m(\eta)$ is given by  
\begin{equation}
m(\eta)  = m_{I}
\sqrt{1 + 2 H_I \eta}, 
\label{m_t}
\end{equation}
where 
$H_I$ is initial data at the  matter production instant.  

The kinetic equation transformed into convenient form for numerical calculations is given in  Appendix.  
In Fig. 1 the evolution of the fermion number density~($g_f$ is the degeneracy factor)
\begin{equation}
n = 2 g_f \int \frac{d^3p}{(2 \pi)^3} f(\vec{p},t)
\label{number}
\end{equation}
for different ratios $\frac{m_{I}}{2 H_I}$ is shown.   
The final spectra  of produced
particles are far from exponential as  seen from Fig. 2.  
The time evolution of specific entropy $S/n$ is shown in Fig. 3, where  entropy $S$ is given by 
\begin{equation}
S = 2 g_f \int \frac{d^3p}{(2 \pi)^3} \left[  f(\vec{p},t) \ln  f(\vec{p},t) + (1-  f(\vec{p},t)) \ln(1- f(\vec{p},t))\right].
\label{ent}
\end{equation}

\subsection{Vector interaction}
In this section we consider some application of the model to the interaction 
of the type $\Delta L = e A_\mu  \bar{\psi} \gamma^\mu \psi$.

\subsubsection{Particle production in the external field with fixed direction}
Here we consider another simple example,   particle production in the external time-dependent 
field with fixed direction. As in the previous subsection,  let us demand zero initial conditions for the particle fields
in the asymptotic past.

The vector potential is given by time-dependent amplitude and constant vector $\vec{n}$:   
\begin{equation}
\vec{A}  =  A (t) \vec{n}.
\label{flux_tube_geometry}
\end{equation}

The KE in this case also can be reduced\footnote{We leave the proof  to the reader.} to the system of three 
equations
\begin{eqnarray}
\label{KE_f_begin}
\dot{f}_0 &=&  2 W v^+, \\
\dot{v}^+ &=&  W (1 - 2 f_0) -  2 \omega v^-, \\
\dot{v}^- &=&  2 \omega v^+, 
\label{KE_f}
\end{eqnarray}
with   $W$, $v^\pm$ defined by 
\begin{eqnarray}
W  &=& |\vec{\Xi}| =  \frac{e |\vec{E}| \sqrt{p_\perp^2+m^2}}{2 \omega^2}, \\
v^\pm &=& \mp (\frac{\Xi_\|}{|\vec{\Xi}|} r^\pm_{\|} + \frac{\Xi_\perp}{|\vec{\Xi}|} r^\pm_{\perp}), 
\label{W_field}
\end{eqnarray}
where the  vector decomposition was used:
\begin{eqnarray}
\vec{P} &=& P_{\|} \vec{n} + \frac{\vec{p}_\perp}{|\vec{p}_\perp|} p_\perp,\\
\vec{r}^{\pm} &=& r^\pm_{\|} \vec{n} + \frac{\vec{p}_\perp}{|\vec{p}_\perp|} r^\pm_\perp.
\label{}
\end{eqnarray}

The set of equations (\ref{KE_f_begin}-\ref{KE_f})  coincides with pioneering
results obtained by Bialynicki-Birula, Gornicki, and Rafelski~\cite{Bialynicki-Birula:1991tx} 
in the framework of the equal-time Wigner function. 
Also, for  the one-component external
field $\vec{A}=(0,0,A_3(t))$, 
the derivation of this KE in the framework of Bogoliubov 
quasiparticles can be found in~\cite{grib80,Smolyansky:1997fc}; some applications 
to particle production in strong laser fields~\cite{laser2,laser1} 
as well as 
in the ultrarelativistic heavy-ion collision~\cite{Prozorkevich:2004yp,Skokov:2004zh}  in the Abelian 
dominance approximation~\cite{Gyulassy} were also considered. 
\label{sec:flux}

\section{Summary}

In the present paper, we use the oscillator  representation~\cite{holomorphic,ps}
to derive  the most general kinetic equation for fermions 
in time-dependent
external electro-magnetic field of arbitrary direction.
Also particle production due to  time dependence of fermion mass 
(as a result of Yukawa-like  interactions in the meanfield approximation
or conformally static  space-time)
is considered.

The derived KE in the  special cases (time-dependent mass, fixed direction of 
the external field) coincides with KE obtained by time-dependent 
Bogoliubov transformation approach.  
However for the latter  there  is an unsolved problem of particle production 
in the vector fields  of  alterating  direction. As  shown in the present paper, 
the oscillator representation  overcomes this difficulty and helps us to derive  KE 
for fermions created in field of general polarization $\vec{A} = (A_1(t),A_2(t),A_3(t))$.

Some applications of derived KE are discussed. 
The detailed investigation and numerical calculations 
are done for fermion  creation in conformal cosmology. 
The solution of KE in this case has shown that the obtained 
momentum distribution function is far from equilibrium, therefore can be 
an argument in favour  that  the introduction of temperature 
is impossible without taking into account collision processes  in created fermion gas.

The consistent description of particle creation in the strong laser fields 
includes the effects of the field direction   alteration. So far as the simplest model 
of the laser field only the linear  polarised one  is 
considered (e.g. ~\cite{Popov:2001ak,Blaschke:2004hb,Blaschke:2005hs,laser2}).
Our approach based on quantum field theory  
allows numerical calculation of particle production in  the field of arbitrary polarisation.  
In this case  the simplifications assumed in  Section 4  are no longer applicable  and one has to deal 
with the general system of equation~(\ref{general_final_beg_}-\ref{general_final_}). 
Preliminary numerical  results on this subject are promising but 
out of the scope of the current paper and will be printed elsewhere. 

\section*{Acknowledgements}
We thank  A. Prozorkevich and S. Smolyansky  for  the fruitful discussions.
One of us (SV) is  also grateful for continuous and
stimulating conversations  with D.~Blaschke,  P.~Levai and   V.~Toneev.

This work is supported in part by MTA-JINR grant, 
the Russian Foundation for Basic Research, grant 
05-02-17695 and a special program of the Ministry of Education and Science of the Russian Federation, 
grant RNP.2.1.1.5409.

\section*{Appendix}
We introduce dimensionless variables of the time $\tau$, momentum $q$ and the parameter 
$\gamma_I$ according to the following formulas: 
\begin{eqnarray}
\tau &=& 2 H_I \eta, \\
\vec{q} &=& \vec{p} / m_{I}, \\
\gamma_I &=& \frac{m_{I}}{2 H_I}. 
\label{var}
\end{eqnarray}

The kinetic equation (\ref{KE_mass_time_dependent_begin}-\ref{KE_mass_time_dependent})
in this variables takes the form
\begin{eqnarray}
\frac{d}{d \tau} f_0  &=& 2 \tilde{W} v^+, \\
\frac{d}{d \tau} v^+  &=& \tilde{W} (1 - 2 f_0) - 2 \tilde{\omega} \gamma_I v^- , \\
\frac{d}{d \tau} v^-  &=& 2 \tilde{\omega} \gamma_I  v^+, 
\label{KE_dimen}
\end{eqnarray}
where 
\begin{eqnarray}
 \tilde{W}  &=& \frac{1}{4} \frac{\vert \vec{q} \vert}{(\vec{q}^2 + 1 + \tau)\sqrt{1+\tau} }, \\
 \tilde{\omega} &=&  \sqrt{\vec{q}^2 + 1 + \tau}. 
\label{varII}
\end{eqnarray}

\end{document}